\documentclass[preprint,pre,showpacs,preprintnumbers]{revtex4}

\usepackage{graphicx}

\begin{document}

\title{Theoretical and numerical investigations on shapes of planar lipid monolayer domains}

\author{Hao Wu} \email[Email address: ]{wadewizard@gmail.com}
\affiliation{Department of Physics, Beijing Normal University,
Beijing 100875, China}
\author{Z. C. Tu}\email[Email address: ]{tuzc@bnu.edu.cn}
\affiliation{Department of Physics, Beijing Normal University,
Beijing 100875, China}

\date{\today}

\begin{abstract}
Shapes of planar lipid monolayer domains at the air-water interface
are theoretically and numerically investigated by minimizing the
formation energy of the domains which consist of the surface energy,
line tension energy, and dipole electrostatic energy. The shape
equation which describes boundary curves of the domains at
equilibrium state is derived from the first order variation of the
formation energy. A relaxation method is proposed to find the
numerical solutions of the shape equation. The theoretical and
numerical results are in good agreement with previous experimental
observation. Some new shapes not observed in previous experiments
are also obtained, which awaits experimental confirmation in the
future.
\end{abstract}
 \maketitle

\maketitle

\section{INTRODUCTION}
In the last three decades lipid monolayer domains (LMDs) at an
air-water interface have drawn much attention from biochemists and
physicists. Experimental researchers have observed that the lipid
monolayer domains display many kinds of shapes, such as circular
form, S form, dumbbell form, serpentine
form\cite{A:Stine,A:K.Y.C.Lee,A:M.M.Lipp}, torus
form\cite{A:R.M.Weis85,A:H.M.McConnell90}, fan form, multi-leaves
form with spiral
arms\cite{A:K.Y.C.Lee,A:M.M.Lipp,A:R.M.Weis85,A:R.M.Weis84,A:H.E.Gaub,A:Peter
Kruger99,A:Peter Kruger} (also called labyrinthine form) and so on,
which brings a theoretical question: How to understand these shapes?
McConnell {\it et~al.} proposed that the shape of LMDs was a natural
result of competition between the energy due to the line tension of
domain boundary and the dipole electrostatic energy of lipid
molecules\cite{A:H.M.McConnell88,A:H.M.McConnell}. By assuming that
all lipid molecules align along the normal direction of the domain
plane, they also proved that the total dipole electrostatic energy
can be simplified as a double curve integral form along the domain
boundary which will be called as McConnell energy in the following
contents. Following McConnell's seminal work, Iwamoto and Ou-Yang
proposed that besides the above two kinds of energy, the growth of a
lipid domain should cost an additional surface energy due to the
difference in the Gibbs free energy density between outer
(liquid-like disorder) and inner (solid-like order)
phases\cite{A:M.Iwamoto04}. Each observed LMD should minimize its ``
formation energy " (i.e., the sum of these three kinds of
enrgy)\cite{A:M.Iwamoto04}. Based on this formation energy, Iwamoto
{\it et~al.} explained the existence of several kinds of
non-circular domains observed in the
experiments\cite{A:M.Iwamoto06,A:M.Iwamoto08,A:Iwamoto08}.

In fact, an implicit assumption to derive the McConnell energy is
that the interaction between two dipoles is proportional to
$\ell^{-3}$ where $\ell$ is the distance between them. It is this
assumption that leads to the divergence of McConnell energy. As is
well known, this assumption does not hold when $\ell$ is close to
the dipole length(i.e., the length of lipid molecules or the
thickness of LMDs). It is safe to use McConnell energy if the least
distance between dipoles in the LMDs is much smaller than the
thickness of LMDs. In the contrary case, the dipole electrostatic
energy should be dealt with in the other way. Fortunately, when
Langer {\it et~al.} investigated the pattern formation in magnetic
fluid\cite{A:S.A.Langer}, as a byproduct, they proved that the total
dipole electrostatic energy can also be simplified as a double
integral form along the domain boundary which will be called as
Langer energy in the following contents. The only difference between
it and the McConnell energy is in the integrand. As a result, the
Langer energy avoids the divergence of dipole electrostatic energy.
In experiments, the thickness of LMDs is about $2$ nanometers while
the least distance between dipoles in the LMDs might be several
angstroms. Therefore, it is of significance to reconsider the shapes
of LMDs based on the Langer energy. In this paper, the formation
energy of a LMD is defined as the sum of the surface energy, the
energy due to the line tension, and the Langer energy. The shape
equation describing the boundary curve of the LMD is derived from
the variation of the formation energy. We obtain many analytical and
numerical solutions to the shape equation which agree well with the
previous experimental results. We also present several kinds of
shapes which have not been observed in experiments, which awaits the
further conformation in the further experiments.

The rest of this paper is organized as follows: In Sec. II, we
present the formation energy of a LMD based on Langer energy, and
then derive the shape equation of LMD by minimizing the formation
energy. In Sec. III, we describe briefly our algorithm to solve the
shape equation. In Sec. IV, we present many analytical and numerical
solutions to the shape equation, and then compare them with
experimental results. The last section is a brief summary and
prospect.

\section{THEORETICAL MODEL}
\begin{figure}[t!]
\begin{center}
\includegraphics[width=8.5cm]{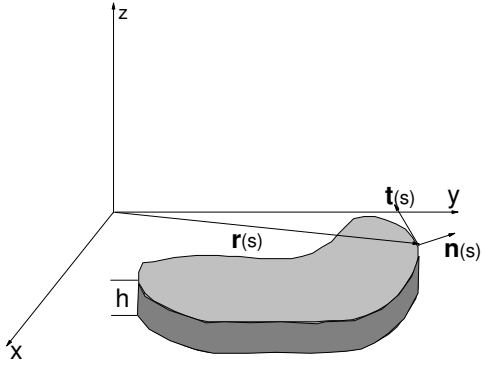}
\caption{\label{F:pd}  Schematic illustration of the slab geometry
of a lipid domain, where $h$, $\mathbf{r}(s)$, $\mathbf{t}(s)$,
$\mathbf{n}(s)$ are respectively the thickness of the domain, the
displacement vector of any point in the boundary curve, the tangent
and normal vectors of the point in the boundary curve. }
\end{center}
\end{figure}

A LMD with thickness $h$ is schematically depicted in
Fig.~\ref{F:pd} where $z-$axis is the normal direction of the LMD
and the projection of LMD in $xy-$plane is enclosed by a boundary
curve $\{\mathbf{r}(s)\}$ with arc-length parameter $s$.
$\mathbf{t}(s)$ and $\mathbf{n}(s)$ represent the tangent and normal
vectors of point $\mathbf{r}(s)$ in the boundary curve,
respectively. They satisfy $d\mathbf{t}/{ds}=\kappa\mathbf{n}$ and
$d\mathbf{n}/{ds}=-\kappa\mathbf{t}$, where $\kappa$ is the
curvature of the boundary curve. For simplicity, we introduce
notations $\mathbf{R}_{ls}=\mathbf{r}(l)-\mathbf{r}(s)$,
$R=|\mathbf{R}_{ls}|$, and
$\hat{\mathbf{R}}_{ls}=\mathbf{R}_{ls}/R$. The Langer energy (i.e.,
the dipole electrostatic energy) can be expressed as
\cite{A:S.A.Langer}:
\begin{equation}
\label{Langerenergy}
F_{L}=\frac{\mu^2}{2h^2}\int_0^hdz\int_0^hdz'\oint_Lds\oint_Ldl\frac{{\bf
t(l)}\cdot{\bf t(s)}}{\sqrt{R^2+(z-z')^2}},
\end{equation}
where $\oint$ represents the integral along the boundary curve.
$\mu$ is the dipole density in the LMD.

The formation energy of the LMD can be expressed as
\begin{equation}
\label{energy} F=\Delta P A + \gamma \oint ds+F_{L},
\end{equation}
where $\Delta P$ is the surface energy density introduced by Iwamoto
and Ou-Yang in Ref.\cite{A:M.Iwamoto04}. $A$ is the surface area of
LMD and $\gamma$ is line tension of the boundary.

Now we will drive the shape equation by using the variational method
proposed in Ref.~\cite{A:Z.C.Tu}. The variation in the tangent
direction of the boundary curve will give a trivial identity. If we
take the variation in the normal direction as
$\delta\mathbf{r(s)}=\Omega(s)\mathbf{n}(s)$, it is not hard to
obtain the following variational formulas:
\begin{eqnarray}
\delta A&=&\oint ds\Omega(s),\label{variationa}\\
\delta \oint(ds)&=&-\oint ds\kappa(s)\Omega(s),\label{variationds}\\
\delta\oint ds\oint dl
\frac{{\mathbf{t}(l)}\cdot{\mathbf{t}(s)}}{\sqrt{R^2+(z-z')^2}} &=&2
 \oint ds\Omega(s) \oint dl
\frac{\mathbf{R}_{ls}\times\mathbf{t}(l)}
{[R^2+(z-z')^2]^{3/2}},\label{variationoo}
\end{eqnarray}
where `` $\times$ " is the 2-dimensional cross product which
satisfies $\mathbf{a}\times\mathbf{b} = a_{1}b_{2}-a_{2}b_{1}$ for
any 2-dimensional vectors $\mathbf{a}\equiv(a_{1},a_{2})$ and
$\mathbf{b}\equiv(b_{1},b_{2})$. By considering
Eqs.(\ref{variationa})-(\ref{variationoo}), and $\int_0^h dz\int_0^h
dz' {[R^2+(z-z')^2]^{-3/2}}=R^{-2}(\sqrt{R^2+h^2}-R)$, we can derive
the shape equation \begin{equation} \label{integ} \Delta
P-\gamma\kappa(s)-(2\mu^2/{h}^2)\oint dl
[\hat{\mathbf{R}}_{ls}\times\mathbf{t}(l)][\sqrt{1+{({h}/{R})}^2}-1]=0
\end{equation} from $\delta F=0$.

\section{NUMERICAL METHOD}
In the last section, we have derive the shape equation (\ref{integ})
of LMDs by using variational method. However, it is very hard to
obtain its solutions. Here we will develop a numerical method to
solve it. In terms of the physical meaning of variational method, we
can express the force in the boundary curve in a vector form as
\begin{equation}\mathbf{f}(s)=-\left\{\Delta
P-\gamma\kappa(s)-(2\mu^2/{h}^2)\oint dl
[\hat{\mathbf{R}}_{ls}\times\mathbf{t}(l)][\sqrt{1+{({h}/{R})}^2}-1]\right\}
\mathbf{n}.\label{forcen}\end{equation} Thus Eq.(\ref{integ})
reflects the force balance ($\mathbf{f}=0$) in the boundary curve.

We introduce a virtual damping system evolving with dynamical
equation \begin{equation}\frac{\partial \mathbf{r}(s,\tau)}{\partial
\tau}=\mathbf{f}(s),\label{relaxeq}\end{equation} where $\tau$ is
virtual time. If a LMD whose boundary curve satisfies
Eq.(\ref{integ}), its shape will keep unchange. On the contrary,
$\mathbf{f}\neq 0$, the boundary curve will evolve according to
Eq.(\ref{relaxeq}). In a long enough time, a damping system can
usually approach to a state satisfying $\partial\mathbf{r}/{\partial
\tau}=0$ which implies $\mathbf{f}=0$. Thus the final configuration
is naturally a solution to the shape equation (\ref{integ}). The
above scheme to find numerical solutions of Eq.(\ref{integ}) is
called the relaxation method, which had also been used to resolve
hydrodynamics of the ferrofluid drop pattern
formation\cite{A:S.A.Langer,A:A.J.Dickstein,A:D.P.Jackson}, shape
relaxation in Langmuir layers\cite{A:R.E.Goldstein} and
hydrodynamics of the monolayer problem at the air-water
interface\cite{A:D.K.Lubensky} by Goldstein and his collaborators.

In order to perform numerical calculations, we should transform the
continuous equation (\ref{relaxeq}) into the discrete form by
dividing the boundary curve into $N$ segments with $N$ points and
assuming the step of evolution time is $\tau_0$. Then we label $N$
points in the boundary as $1,2,\cdots,k-1,k,k+1,\cdots,N$ and the
time sequence as
$\tau_0,2\tau_0,\cdots,(j-1)\tau_0,j\tau_0,(j+1)\tau_0,\cdots$.
Eq.(\ref{relaxeq}) is transformed into
\begin{equation}\mathbf{r}_{j+1}^k-\mathbf{r}_{j}^k=\mathbf{f}_j^k \tau_0,\label{disrelaxeq}\end{equation}
where $\mathbf{f}_j^k$ is the discrete form of Eq.(\ref{forcen}).
The geometric quantities in this equation can be discreted as:
\begin{eqnarray}
[ds]_j^k&=&|\mathbf{r}_{j}^{k+1}-\mathbf{r}_{j}^k|,
\end{eqnarray}
\begin{eqnarray}
[\mathbf{t}]_j^k&=&\frac{\mathbf{r}_{j}^{k+1}-\mathbf{r}_{j}^k}{[ds]_j^k},
\end{eqnarray}
\begin{eqnarray}
[\kappa\
\mathbf{n}]_j^k&=&\frac{\mathbf{t}_{j}^{k+1}-\mathbf{t}_{j}^k}{[ds]_j^k},
\end{eqnarray}
where $\mathbf{n}_j^k$ can be obtained by rotating $\mathbf{t}_j^k$
in $90^\circ$ clockwise.

The stop condition of our calculation of Eq.(\ref{disrelaxeq}) is
$\sum_{k=1}^N|\mathbf{r}_{J+1}^k-\mathbf{r}_{J}^k|<\epsilon$ for a
large enough integer $J$ and a small enough number $\epsilon$.

\section{RESULTS}

\subsection{ANALYTICAL SOLUTIONS}

\subsubsection{Circular solutions}

For a circle with radius $R_{0}$, $\kappa=-\frac{1}{R_{0}}$. Because
${\hat{\bf R}}_{ls}$ and ${\bf t}(l)$ are unit vectors, we can
respectively express them as $(\cos\theta_{ls}, \sin\theta_{ls},0)$
and $(\cos\theta_{l}, \sin\theta_{l},0)$. Eq.~(\ref{integ}) is then
transformed into:
\begin{eqnarray}
\label{ellip} ax+b=xE(2ix)-2x^2
\end{eqnarray}
where $x=\frac{R_{0}}{h}$, $a=\frac{\Delta P h}{4\mu^2}$ ,
$b=\frac{\gamma}{4\mu^2}$, and $E(.)$ is complete elliptic integrals
of the second kind.

\begin{figure}[t!]
\begin{center}
\includegraphics[width=8.5cm]{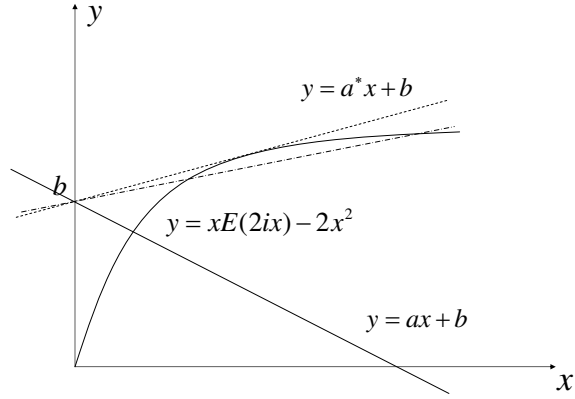}
\caption{\label{F:so}  Schematic diagrams of functions $y=ax+b$ and
$y=xE(2ix)-2x^2$. A critical $a^{\ast}$ exists so that there are two
solutions when $0<a<a^{\ast}$ and no solution when $a>a^{\ast}$.
Additionally, there is only one solution for $a<0$ or $a=a^{\ast}$.
}
\end{center}
\end{figure}

From Fig \ref{F:so}, we find that a critical $a^{\ast}$ exists such
that there are two solutions when $0<a<a^{\ast}$ and no solution
when $a>a^{\ast}$. Additionally, there is only one solution for
$a<0$ or $a=a^{\ast}$.

In the limit case of $x\rightarrow\infty$, we have
\begin{eqnarray}
\label{ellip3} x E(2ix)-2x^2\approx\frac{1}{4}\ln(8{\sqrt e}x),
\end{eqnarray}
from which we easily obtain the radius of the circular domain
\begin{eqnarray}
\label{sol1} R_{0}=\frac{e^{\frac{\gamma}{\mu^2}}}{8\sqrt e}h,
\end{eqnarray}
for vanishing $\Delta P$.

This result is similar to McConnell's result
\begin{eqnarray}
R_{0}=\frac{e^3e^{\frac{\gamma}{\mu}}}{4}\delta
\end{eqnarray}
with a cutoff $\delta$ \cite{A:H.M.McConnell88}.

\subsubsection{Toroidal solutions}

Next, we investigate whether toroidal shapes are permitted in our
model. The torus solution is mentioned in many
references\cite{A:R.M.Weis85,A:H.M.McConnell90,A:M.Iwamoto04,A:M.Iwamoto06}.
Here we only discuss the case of $h\ll min\{R_{1}-R_{2},R_{2}\}$,
where $R_{1}$ and $R_{2}$ are the outer and inner radii,
respectively.

From Eq.~(\ref{integ}) we find the radii $R_{1}$ and $R_{2}$ should
satisfy
\begin{eqnarray}
\label{outer} \Delta P
R_{1}+\gamma-\mu^2\ln\frac{8\sqrt{e}R_{1}}{h}+\mu^2\int_0^{2\pi}\frac{x(x-\cos\theta)d\theta}{(x^2+1-2x\cos\theta)^{\frac{3}{2}}}dl=0,
\end{eqnarray}
and
\begin{eqnarray}
\label{inner} \Delta P
R_{2}-\gamma+\mu^2\ln\frac{8\sqrt{e}R_{2}}{h}-\mu^2\int_0^{2\pi}\frac{x(1-x\cos\theta)d\theta}{(x^2+1-2x\cos\theta)^{\frac{3}{2}}}dl=0,
\end{eqnarray}
where $x=R_{2}/R_{1}$.

\begin{figure}[t!]
\begin{center}
\includegraphics[width=8.5cm]{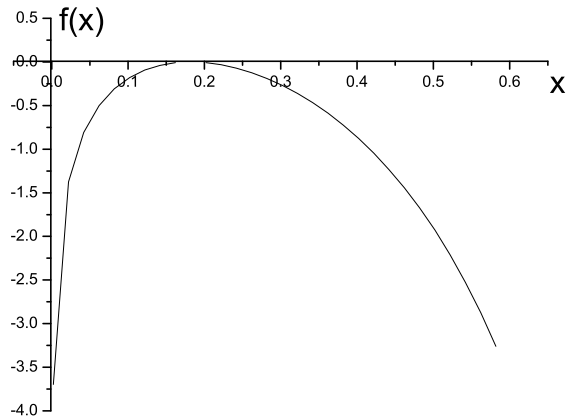}
\caption{\label{F:hp}  Schematic diagram of function
$f(x)=\alpha(1+x)+\ln x-2[E(k)-K(k)]$ when $\alpha\approx2.525115$,
there is a solution $x=R_{2}/R_{1}\approx0.182215$. }
\end{center}
\end{figure}

By adding Eq.~(\ref{outer}) to Eq.~(\ref{inner}), we obtain
\begin{eqnarray}
\label{ell2} \alpha(1+x)+\ln x-2[E(k)-K(k)]=0
\end{eqnarray}
where $\alpha=\frac{\Delta P R_{1}}{\mu^2}$, $k^2=-4x/(1-x)^2$.
$K(k)$ and $E(k)$ are respectively complete elliptic integrals of
the first kind and the second kind, respectively. As shown in Fig.
\ref{F:hp}, there is a critical solution $x\approx 0.182215$ when
$\alpha\approx 2.525115$. This solution is close to the ratio $0.2$
observed in the experiment \cite{A:H.M.McConnell90}. For
$\alpha>2.525115$ there are two toroidal solutions.

Therefore toroidal shapes are indeed permitted under certain
conditions in our model.

\subsection{NUMERICAL SOLUTIONS}

There are four parameters, $\mu^2$ , $\gamma$ , $h$ and $\Delta P$
in our model. We take the experimental values $h\approx2nm$,
$\mu^2=5\times10^{-9}dyn$, and $\gamma=1.6\times10^{-7}dyn$
\cite{A:D.J.Benvegnu}. The surface energy density $\Delta P$ is an
adjustable parameter, which can be understood as the Gibbs free
energy difference between the fluid phase and solid phase
\cite{A:M.Iwamoto04}.

To make sure of our numerical method, we consider the circular
domains. The radius of circular domains is about $20\mu m$ in the
experiment\cite{A:D.J.Benvegnu}. Using the theoretical result
Eq.~(\ref{ellip3}), we obtain $\Delta
P\approx-5.05255\times10^{-5}dyn/cm$. Adopting this parameter in our
simulations, we obtain stable circular domain with radius about
$20\mu m$.

\begin{figure}[t!]
\begin{center}
\includegraphics[width=8.5cm]{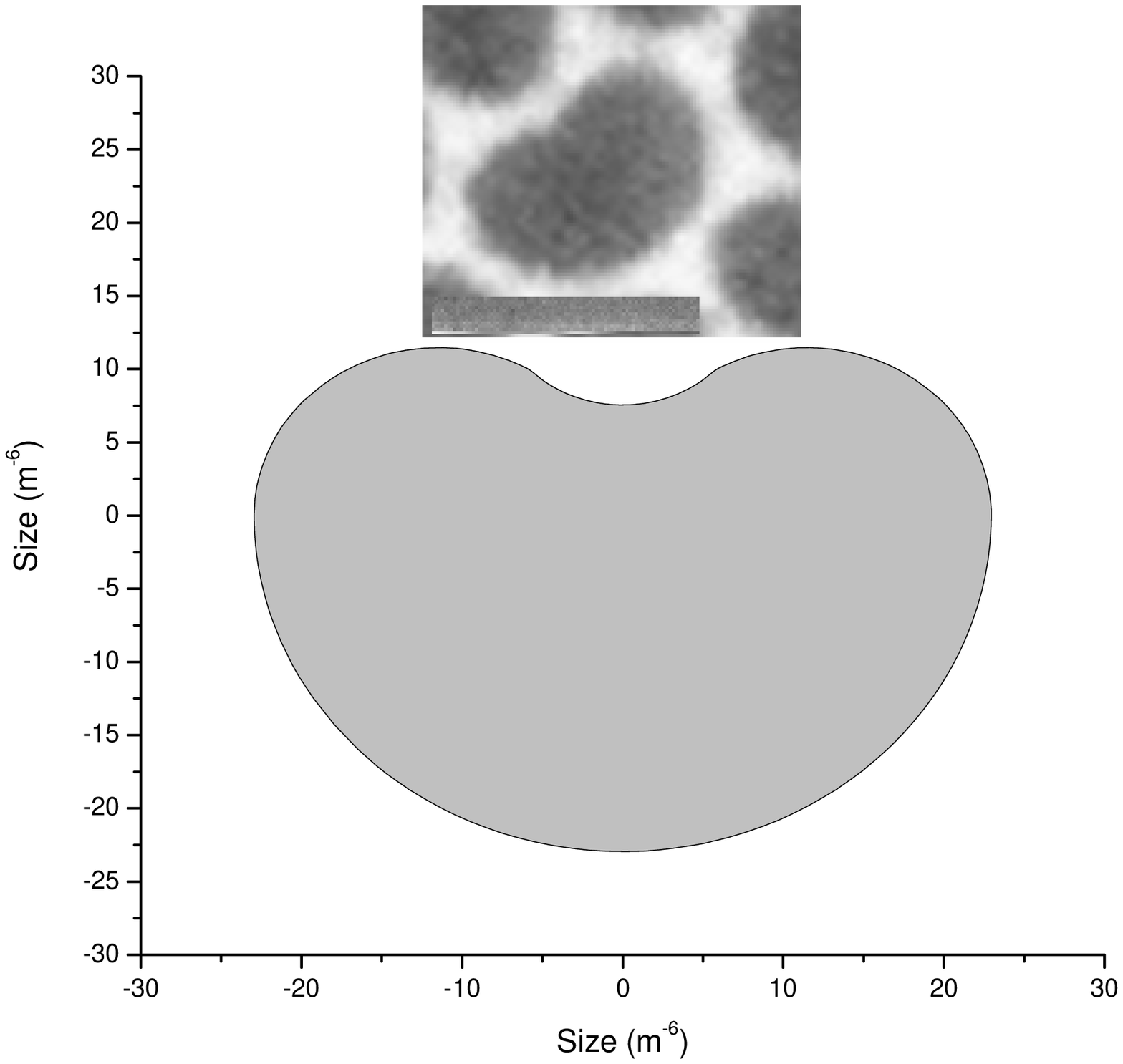}
\caption{\label{F:hna}  Smooth heart. The top figure is cut from
Fig.5 (B) in Ref.\cite{A:Stine}, where the bottom one is the
numerical result with adjustable parameter $\Delta
P\approx1.19338\times10^{-4}dyn/cm$. }
\end{center}
\end{figure}

\begin{figure}[t!]
\begin{center}
\includegraphics[width=8.5cm]{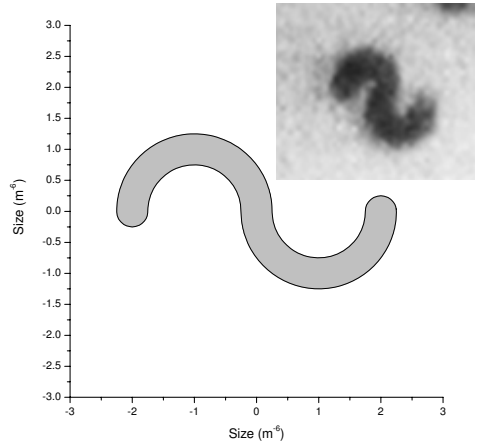}
\caption{\label{F:S2}  S domain. The figure in the top right hand
corner is cut from the second picture of Fig.3 in
Ref.\cite{A:H.E.Gaub}, where the bottom left one is the numerical
result with adjustable parameter $\Delta
P\approx3.87582\times10^{-3}dyn/cm$. }
\end{center}
\end{figure}

\begin{figure}[t!]
\begin{center}
\includegraphics[width=8.5cm]{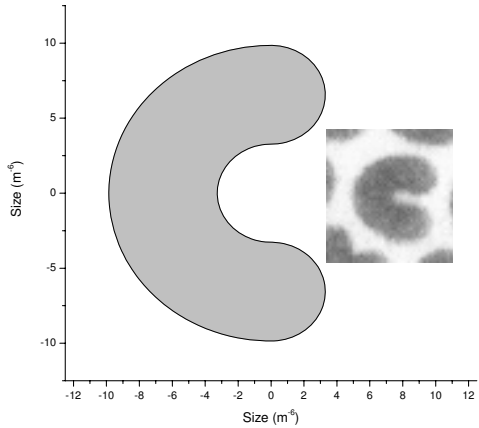}
\caption{\label{F:yueya}  Moon. The right figure is cut from Fig.5
(D) in Ref.\cite{A:Stine}, where the left one is the numerical
result with adjustable parameter $\Delta
P\approx-1.42196\times10^{-4}dyn/cm$. }
\end{center}
\end{figure}

\begin{figure}[t!]
\begin{center}
\includegraphics[width=8.5cm]{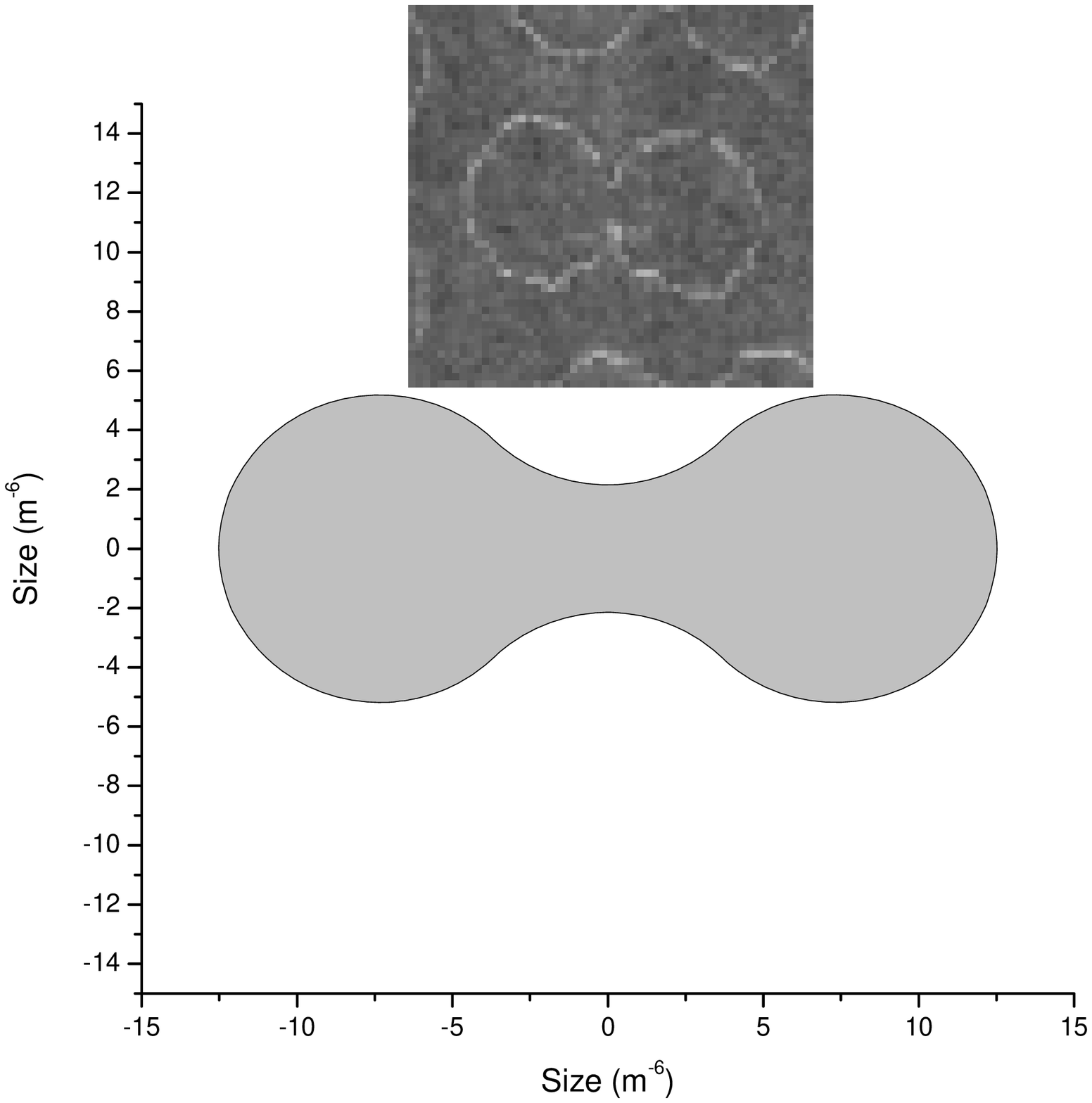}
\caption{\label{F:yaling}  Dumbbell. The top figure is cut from
Fig.1 in Ref.\cite{A:Stine}, where the bottom one is the numerical
result with adjustable parameter $\Delta
P\approx2.01274\times10^{-4}dyn/cm$. }
\end{center}
\end{figure}

Now we use our numerical codes to search for various shapes of
domains evolving from different initial configurations and $\Delta
P$, and then compare them with the experimental results. After the
long time calculations, we obtain heart form (Fig.\ref{F:hna}), S
form (Fig.\ref{F:S2}), moon form (Fig.\ref{F:yueya}) and dumbbell
form (Fig.\ref{F:yaling}) with adjustable parameters $\Delta
P\approx1.19338\times10^{-4}dyn/cm$, $3.87582\times10^{-3}dyn/cm$,
$-1.42196\times10^{-4}dyn/cm$, and $2.01274\times10^{-4}dyn/cm$,
respectively. These shapes and their sized are in good agreement
with the experiments\cite{A:Stine,A:H.E.Gaub}.

\begin{figure}[t!]
\begin{center}
\includegraphics[width=8.5cm]{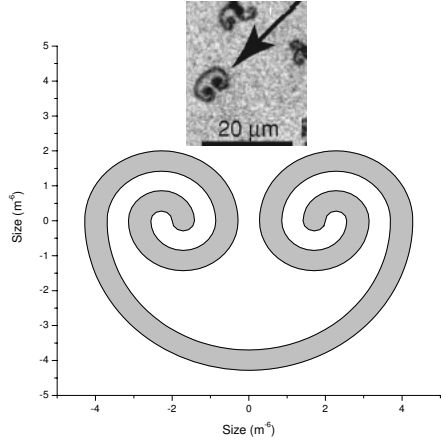}
\caption{\label{F:GS}  One-lobed domain. The top figure is cut from
Fig.4 in Ref.\cite{A:Peter Kruger}, where the bottom one is the
numerical result with adjustable parameter $\Delta
P\approx-3.97255\times10^{-3}dyn/cm$. }
\end{center}
\end{figure}
\begin{figure}[t!]
\begin{center}
\includegraphics[width=8.5cm]{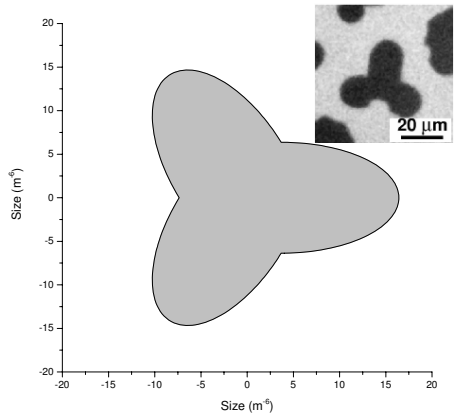}
\caption{\label{F:f3}  Three-leaves form without spiral arms. The
figure in the top right hand corner is cut from the fifth picture of
Fig.3 in Ref.\cite{A:Peter Kruger}, where the bottom left one is the
numerical result with adjustable parameter $\Delta
P\approx-1.66338\times10^{-4}dyn/cm$. }
\end{center}
\end{figure}
\begin{figure}[t!]
\begin{center}
\includegraphics[width=8.5cm]{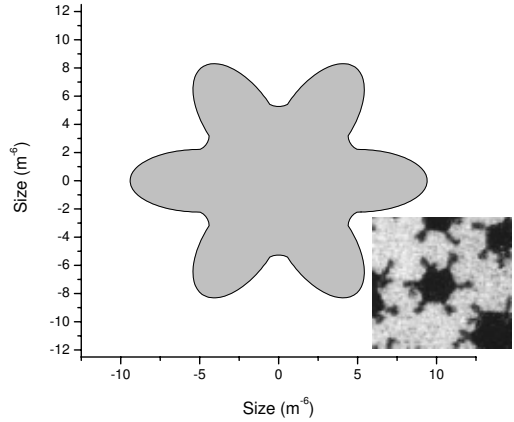}
\caption{\label{F:f6}  Smooth six-leaves form without spiral arms.
The figure in the bottom right hand corner is cut from the second
picture of Fig.3 in Ref.\cite{A:Peter Kruger99}, where the top left
one is the numerical result with adjustable parameter $\Delta
P\approx-3.11274\times10^{-4}dyn/cm$.}
\end{center}
\end{figure}
\begin{figure}[t!]
\begin{center}
\includegraphics[width=8.5cm]{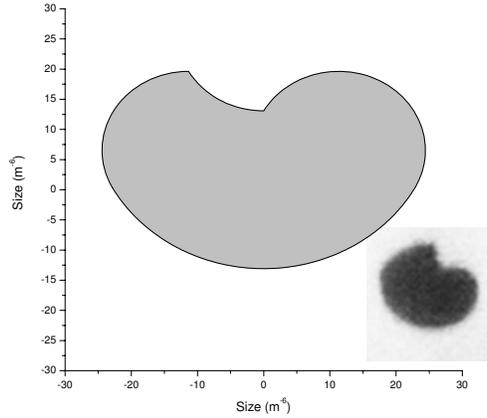}
\caption{\label{F:bean}  Bean with two cusps. The top figure in the
bottom right hand corner is cut from the first picture of Fig.3 in
Ref.\cite{A:H.E.Gaub}, where the top left one is the numerical
result with adjustable parameter $\Delta
P\approx9.94745\times10^{-5}dyn/cm$. }
\end{center}
\end{figure}
\begin{figure}[t!]
\begin{center}
\includegraphics[width=6.5cm]{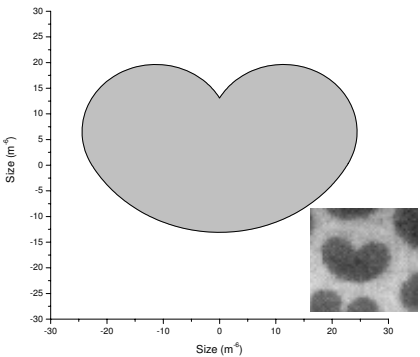}
\caption{\label{F:ha}  Heart with a cusp. The figure in the bottom
right hand corner is cut from Fig.5 (D) in Ref.\cite{A:Stine}, where
the top left one is the numerical result with adjustable parameter
$\Delta P\approx9.72545\times10^{-5}dyn/cm$. }
\end{center}
\end{figure}
\begin{figure}[t!]
\begin{center}
\includegraphics[width=8.5cm]{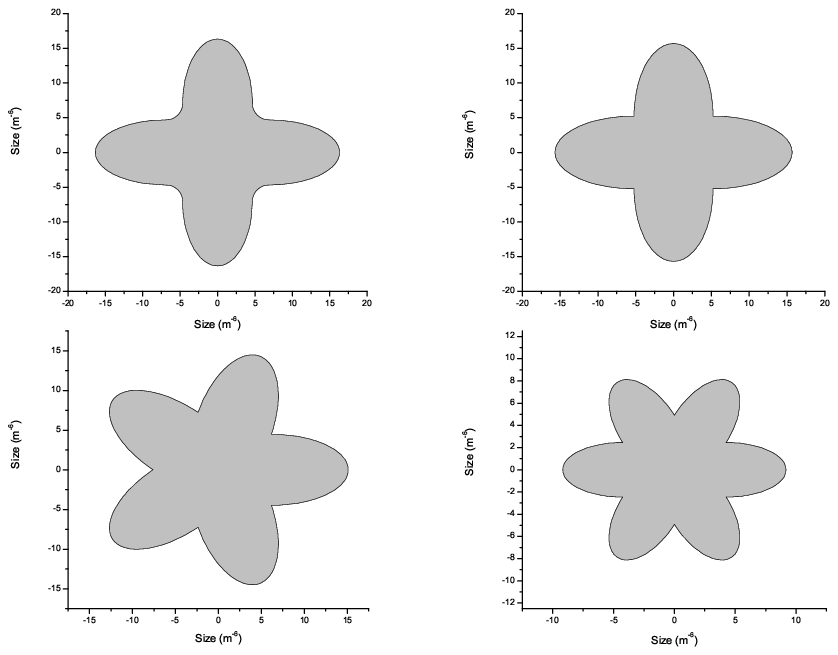}
\caption{\label{F:g1}  Multi-leaves forms. Top left, four-leaves
form without cusps when $\Delta P\approx1.63377\times10^{-5}dyn/cm$.
Top right, four-leaves form with cusps when $\Delta
P\approx-2.06338\times10^{-4}dyn/cm$. Bottom left, five-leaves form
with cusps when $\Delta P\approx-8.63377\times10^{-5}dyn/cm$. Bottom
right, six-leaves form with cusps when $\Delta
P\approx-1.34127\times10^{-3}dyn/cm$. }
\end{center}
\end{figure}

In addition, our model can also obtain one-lobed domain (also called
labyrinthine pattern) and three-leaves domain shown in
Figs.~\ref{F:GS} and \ref{F:f3} under the parameters $\Delta
P\approx-3.97255\times10^{-3}dyn/cm$ and
$-1.66338\times10^{-4}dyn/cm$, respectively. Their shapes and sizes
are similar to those observed in Peter Kr\"{u}ger and Mathias
L\"{o}sche's experiment \cite{A:Peter Kruger}.

Furthermore, we not only obtain a six-leaves form, which occurs in
the experiment\cite{A:Peter Kruger99} in recent years, shown in
Fig.\ref{F:f6} with adjustable parameter $\Delta
P\approx-3.11274\times10^{-4}dyn/cm$, but also acquire some shapes
with cusps observed in the experiments shown in Figs.\ref{F:bean}
and \ref{F:ha} with adjustable parameters $\Delta
P\approx9.94745\times10^{-5}dyn/cm$ and $\Delta
P\approx9.72545\times10^{-5}dyn/cm$, respectively. These results
agree well with the experiments \cite{A:Stine,A:H.E.Gaub}.

Besides the shapes shown above, we can also produce some new stable
shapes such as four-leaves, five-leaves, and six-leaves with cusps
shown in Fig.\ref{F:g1}, which need to be verified by further
biochemical experiments.

The amazing aspect in our results is that we can obtain some stable
shapes with ``cuspidal points''. Here the cusps are apparent ones
viewed in the scale of micrometers. In our computational procedure,
each cusps locally contain many discrete points in the scale of tens
of nanometers such that the cusps are in fact the smooth curves. We
find the average free energy per unit area of domains with cusps is
less than those without cusps, that is, the former is stable than
the latter. We show the bean shape, heart shape and multi-leaves
shapes with cuspidal points in Figs.\ref{F:bean}, \ref{F:ha} and the
last three pictures in Fig.\ref{F:g1}. In the latest theoretical
work \cite{A:M.Iwamoto08,A:Iwamoto08}, Iwamoto {\it et~al}.
analytically prove that the existence of the cuspidal points is
reasonable and there are two kinds of cusps, one's curvature is
positive, the other's is negative. However, their method is a local
theory, in which one equilibrium domain shape can produce only one
type of cusp and the number is also one. Our theory is a global
theory, because we directly calculate the double curve integrals and
do not use any approximate expansion method. Through numerical
simulation, we further find that two kinds of cusps can coexist in
one equilibrium domain, as shown in Fig.\ref{F:bean}, in which there
are two kinds of cusps. In fact, the domains with cusps have clearly
been observed in serval experiments
\cite{A:Stine,A:H.E.Gaub,A:D.K.Schwartz,A:C.W.McConlogue}. Thus the
shapes with cuspidal points might be widespread in nature.

\begin{figure}[t!]
\begin{center}
\includegraphics[width=8.5cm]{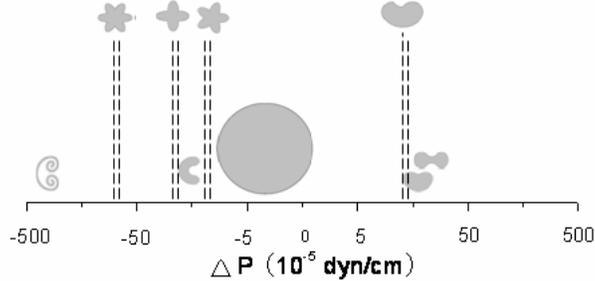}
\caption{\label{F:itg} Phase diagram with various $\Delta P$ (in the
Log coordinate). }
\end{center}
\end{figure}

We have investigated various shapes of domains under different
$\Delta P$. Now we draw a phase diagram in Fig.~\ref{F:itg}. It is
found that the domain shape will become complex from the original
circular domain when the parameter $\Delta P$ changes from
$-5.05255\times10^{-5}dyn/cm$ towards both ends of the coordinate
axis. It seems that the domains with cusps generally occur in the
narrow ranges of parameter $\Delta P$. Two shapes of domains can
coexist in a certain interval, such as heart shape and dumbbell
shape as shown in Figs.\ref{F:itg}, which agrees with the
experimental observations\cite{A:Stine,A:H.E.Gaub} that reveals some
shapes can coexist under some parameters.

\section{CONCLUSION}

In the above discussions, we have theoretically and numerically
investigated the shapes of planar lipid monolayer domains by
minimizing the formation energy of the domains. We obtain some
shapes, such as circular form, S form, dumbbell form, serpentine
form and achiral multi-leaves form and so on. Most of them are
observed in the experiments \cite{A:Stine,A:H.E.Gaub,A:Peter
Kruger99,A:Peter Kruger}. The four-leaves, five-leaves, and
six-leaves forms await the future experimental conformations.

Our model might be extended to explain the shapes of rafts
\cite{A:K.Simons} consisting of sphingolipids and cholesterol in the
cell membranes. Although the cell membrane is a bilayer, the raft
domains are generally in one of the monolayers. There are net
dipoles in the raft domains, thus our model might be applied to
describe the shape of raft domains.

\section*{Acknowledgments}

The authors acknowledge a grant from the Nature Science Foundation
of China (Grant No.10704009), and H.W. acknowledges Z.G. Zheng
(Beijing Normal University) for his support of the research. H.W. is
grateful to Z.C. Ou-Yang (Chinese Academy of Sciences), Y.-J. Wang
(Beijing Normal University), X.T. Wu (Beijing Normal University), F.
Liu (Tsinghua University) for their useful suggestions and kind
help. H.W. also thanks H.M. McConnell (Stanford University) for his
kind reply.


\begin{thebibliography}{10}

\bibitem{A:Stine}
K.J. Stine, and D.T. Stratmann, Langmuir {\bf 8}, 2509 (1992).

\bibitem{A:K.Y.C.Lee}
K.Y.C. Lee and H.M. McConnell, J. Phys. Chem. {\bf 97}, 9532 (1993).

\bibitem{A:M.M.Lipp}
M.M. Lipp, K.Y.C. Lee, J.A. Zasadzinski, and A.J. Waring, Science
{\bf 273}, 1196 (1996).

\bibitem{A:R.M.Weis85}
R.M. Weis and H.M. McConnell, J. Phys. Chem. {\bf 89}, 4453 (1985).

\bibitem{A:H.M.McConnell90}
H.M. McConnell, P.A. Rice, and D.J. Benvegnu, J. Phys. Chem. {\bf
94}, 8965 (1990).

\bibitem{A:R.M.Weis84}
R.M. Weis and H.M. McConnell, Nature {\bf 310}, 47 (1984).

\bibitem{A:H.E.Gaub}
H.E.Gaub, V.T. Moy, and H.M. McConnell, J. Phys. Chem. {\bf 90},
1721 (1986).

\bibitem{A:Peter Kruger99}
P. Kr\"{u}ger, M. Schalke, Z. Wang, R.H. Notter, R.A. Dluhy, and M.
L\"{o}sche, Biophys. J. {\bf 77}, 903 (1999).

\bibitem{A:Peter Kruger}
P. Kr\"{u}ger and M. L\"{o}sche, Phys. Rev. E {\bf 62}, 7031 (2000).

\bibitem{A:H.M.McConnell88}
H.M. McConnell and V.T. Moy, J. Phys. Chem. {\bf 92}, 4520 (1988).

\bibitem{A:H.M.McConnell}
H.M. McConnell, J. Phys. Chem. {\bf 94}, 4728 (1990).

\bibitem{A:M.Iwamoto04}
M. Iwamoto and Z.C. Ou-Yang, Phys. Rev. Lett. {\bf 93}, 206101
(2004).

\bibitem{A:M.Iwamoto06}
M. Iwamoto, F. Liu, and Z.C. Ou-Yang, J. Chem. Phys. {\bf 125},
224701 (2006).

\bibitem{A:M.Iwamoto08}
M. Iwamoto, F. Liu, and Z.C. Ou-Yang, Eur. Phys. J. E {\bf 27}, 81
(2008).

\bibitem{A:Iwamoto08}
M. Iwamoto, F. Liu, and Z.C. Ou-Yang, Int. J. Mod. Phys. B {\bf 22},
2047 (2008).

\bibitem{A:S.A.Langer}
S.A. Langer, R.E. Goldstein, and D.P. Jackson, Phys. Rev. A {\bf
46}, 4894 (1992).

\bibitem{A:Z.C.Tu}
Z.C. Tu and Z.C. Ou-Yang, J. Comput. Theor. Nanosci. {\bf 5}, 422
(2008).

\bibitem{A:A.J.Dickstein}
A.J. Dickstein, S. Erramilli, R.E. Goldstein, D.P. Jackson and S.A.
Langer, Science {\bf 261}, 1012 (1993).

\bibitem{A:D.P.Jackson}
D.P. Jackson, R.E. Goldstein and A.O. Cebers, Phys. Rev. E {\bf 50},
298 (1994).

\bibitem{A:R.E.Goldstein}
R.E. Goldstein and D.P. Jackson, J. Phys. Chem. {\bf 98}, 9626
(1994).

\bibitem{A:D.K.Lubensky}
D.K. Lubensky and R.E. Goldstein, Phys. Fluids. {\bf 8}, 843 (1996).

\bibitem{A:D.J.Benvegnu}
D.J. Benvegnu and H.M. McConnell, J. Phys. Chem. {\bf 96}, 6820
(1992).

\bibitem{A:D.K.Schwartz}
D.K. Schwartz, M.-W. Tsao, and C.M. Knobler, J. Chem. Phys. {\bf
101}, 8258 (1994).

\bibitem{A:C.W.McConlogue}
C.W. McConlogue and T.K. Vanderlick, Langmuir {\bf 13}, 7158 (1997).

\bibitem{A:K.Simons}
K. Simons and E. Ikonen, Nature {\bf 387}, 569 (1997).



\end{thebibliography}
\end{document}